\documentclass[]{aa} 
\usepackage{graphicx}
\usepackage{txfonts}
\usepackage{amssymb}
\usepackage{lineno}
\usepackage{txfonts}
\usepackage{natbib}   
\usepackage{amstext}
\usepackage{mathabx}
\usepackage{multirow}
\usepackage{ulem}
\usepackage{color}
\usepackage{pifont}
\usepackage{lscape}
\usepackage{mathtools}
\usepackage{enumitem}

\usepackage{hyperref}
\begin{document} 


	\title{Free-free absorption parameters of Cassiopeia A\\ from low-frequency interferometric observations}
	\titlerunning{Free-free absorption parameters of Cassiopeia A from low-frequency interferometric observations}


   \author{Lev~A.~Stanislavsky\inst{1}
          \and
          Igor~N.~Bubnov\inst{1}
          \and
          Aleksander~A.~Konovalenko\inst{1}
          \and
					Aleksander~A.~Stanislavsky\inst{1}
					\and
					Serge~N.~Yerin\inst{1,2}
          }

   \institute{Institute of Radio Astronomy, National Academy of Sciences of Ukraine, Mystetstv St., 4, 61002 Kharkiv, Ukraine\\
              \email{lev.stanislavskiy@gmail.com}
          \and
             V.N. Karazin Kharkiv National University, Svobody Sq., 4, 61022 Kharkiv, Ukraine
             }

   \date{Received ; Accepted }

 
  \abstract
   {Cassiopeia A is one of the most extensively studied supernova remnants (SNRs) in our Galaxy. The analysis of its continuum spectrum through low frequency observations plays an important role for understanding the evolution of the radio source and the propagation of synchrotron emission to observers through the SNR environment and the ionized interstellar medium.}
   {In this paper we present measurements of the integrated spectrum of Cas A to characterize the properties of free-free absorption towards this SNR. We also add new measurements to track its slowly evolving and decreasing integrated flux density.}
   {We use the Giant Ukrainian radio telescope (GURT) for measuring the continuum spectrum of Cassiopeia A within the frequency range of 16--72 MHz. The radio flux density of Cassiopeia A relative to the reference source of the radio galaxy Cygnus A has been measured on May--October, 2019 with two sub-arrays of the GURT, used as a two-element correlation interferometer.}
   {We determine magnitudes of emission measure, electron temperature and an average number of charges of the ions for both internal and external absorbing ionized gas towards Cassiopeia A. Generally, their values are comparable to those presented by \citet{Arias2018}, with slight differences. In the absence of clumping we find the unshocked ejecta of $M=2.61M_\sun$ at the electron density of 15.3 cm$^{-3}$ has a gas temperature of $T\approx 100$ K. If the clumping factor is 0.67, then the unshocked ejecta of 0.96$M_\sun$ has the electron density of 18.7 cm$^{-3}$.}
   {The integrated flux density spectrum of Cassiopeia A obtained with the GURT interferometric observations is consistent with the theoretical model within measurement errors and also reasonably consistent with other recent results in the literature.}


   \maketitle

\section{Introduction}
Cassiopeia A (or briefly Cas A) is a very bright historical supernova remnant (SNR) in our Galaxy whose radio emission has been studied for over 75 years \citep{Ryle1948}. This remnant has been under scrutiny like no other. One of the reasons is that Cas A (3C461) is one of the brightest radio sources outside our solar system. Its radiation is often used as a calibration signal to test the response of radio astronomy instruments \citep{Baars1977}. On the other hand, the young core-collapse remnant has a secular (age related due to adiabatic expansion) decrease in the flux density, historically first predicted by \citet{Shklovskii1960} and thereafter detected at 81 MHz \citep{Hogbom1961}. The physical essence of such objects is that shell-type SNRs arise from massive stars that explode. The explosion generates a blast wave that interacts first with the circumstellar medium (CSM) and eventually with the interstellar medium (ISM), generating particle accelerating shocks. At the shock front, charged particles are accelerated in a wide range of energies and emit radio waves, X-rays, and even $\gamma$-rays \citep{Vink2003}. Following the nonthermal (synchrotron) nature, the radio source is brighter at long wavelengths, but its maximum intensity falls at frequencies below 25 MHz due to thermal absorption \citep{Dubner2015}. Low-frequency radio observations give unique information on the evolution of SNRs and their interaction with the ionized ISM along the line of sight that are sensitive to absorption processes \citep{Kassim1995,Gasperin2020}. Moreover, Cas A is one of the few SNRs to exhibit unshocked ejecta through observations \citep{Raymond2018}. Although the ejecta are neutral and very cold early on, they are photoionized and heated by UV and X-ray emission from the ejecta and through the reverse and forward shocks. The radio spectrum of Cas A is significantly modified by internal absorption from the unshocked ejecta. Studying the morphology and spectrum of Cas A at low frequencies, the mass and density in the unshocked ejecta were estimated by \citet[and references therein]{Arias2018}. The external absorption in the ISM also has a noticeable effect on the radio spectrum of Cas A. Therefore, both types of absorption processes should be taken into account for comparison of experimental radio spectra of the SNR with theoretical models at low frequencies. The combination of two electron populations and a time-varying mass in the unshocked ejecta could be responsible for variations superimposed on a smooth secular decline in the flux density of Cas A \citep{Helmboldtl2009,Arias2018}. Unfortunately, this phenomenon remains puzzling, although it has been studied for a long time and continues today \citep{Patnaude2011,Vinyaikin2014,Trotter2017}.

It is important to note that at very low frequencies (< 100 MHz), there is much less available data on the evolution of the Cas A spectrum than at high frequencies. In the low-frequency range, relative methods for measuring the flux density of radio emission are often used, and as a reference radio source in this case, the radio galaxy Cygnus A (briefly Cyg A) is used \citep{Vinyaikin2007,Bubnov2014}. Its flux density spectrum is almost unchanged over time. Moreover, in the low frequency range the radio source (3C405) is closest in magnitude to the source Cas A.

This paper is organized as follows. Starting with important features of our instrumental support, we describe radio observations carried out during May--October, 2019. Our measurements of the Cas A flux density are being made relative to Cyg A used as a reference source. This allows us to compare the experimental spectrum of Cas A at 16--72 MHz with its model accounting for the free-free absorption effects in the ISM and the unshocked ejecta within the SNR. Knowing the flux-density spectrum of Cas A in 2019, we find its spectral index, measure parameters of absorbing regions and estimate the secular decline of the radio flux emitted by Cas A. Finally, we discuss possible reasons of these effects and summarize our findings.

\section{Observations and Facilities}

\subsection{GURT radio telescope configured as a two element interferometer}\label{section2_1}
\begin{figure}
\centerline{\includegraphics[width=0.7\hsize]{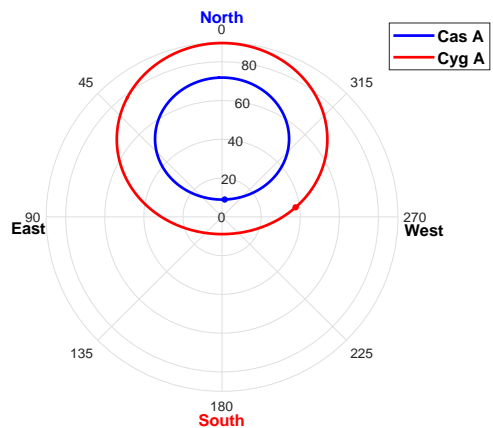}}
\caption{\label{fig1_proj}
Paths of Cas A and Cyg A on the sky as observed from the GURT over 24 hours. The marker on both paths shows a position of each source at 00:01:00 UT on August 8, 2019. This plot displays the position in zenith angle
(90$^\circ -$ altitude) and azimuth.}
\end{figure}

\begin{figure}
\centerline{\includegraphics[width=0.7\hsize]{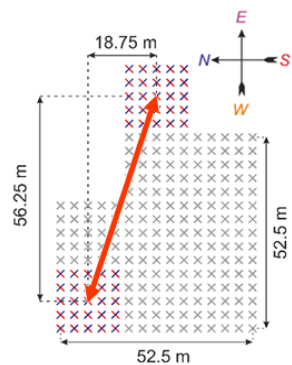}}
\caption{\label{fig2_base}
Location of GURT sub-arrays used in the experiments. The two-element interferometer baseline is noncoplanar. The GURT array consists of 225 active cross-dipoles shown by crosses.}
\end{figure}

The geographical location (49.6$^\circ$N) of the Giant Ukrainian radio telescope (GURT) allows observations of Cas A and Cyg A sources at almost the same zenith angle $\approx$9$^\circ$ at the time of their upper culminations (Figure~\ref{fig1_proj}). The instrument is located near Kharkiv (Ukraine). The GURT is intended for receiving radio emission within the frequency range of 8--80 MHz \citep{Konovalenko2016}. The observations were carried out with two sub-arrays each of which consists of 5 $\times$ 5 active cross-dipoles. The distance between their phase centers is about 60 m. The two-element correlation interferometer has a non-coplanar baseline (Figure~\ref{fig2_base}). 

The effective area of the radio telescope section (5 $\times$ 5 elements) in the direction of zenith (with a deviation of up to $\sim$10$^\circ$), when the distance between the elements $\leq\lambda/2$, takes the form
\begin{equation}
A_{eff}=\left(d(n-1)+\lambda/2\right)^2 = b^2\,,\label{eq1}
\end{equation}
where $d$ is the distance between elements, $n$ the number of elements in a row, $\lambda$ the wavelength. At 40 MHz the value equals to $\approx$ 350 m$^2$. The power pattern of the sub-array reads
\begin{equation}
\left(E(u,v)\right)^2=\left(\frac{\sin\left(\frac{\pi b}{\lambda}(u-u_i)\right)}{\frac{\pi b}{\lambda}(u-u_i)}\times\frac{\sin\left(\frac{\pi b}{\lambda}(v-v_i)\right)}{\frac{\pi b}{\lambda}(v-v_i)}\right)^2\,,\label{eq2}
\end{equation}
where $u$ and $v$ are direction cosines ($u=\cos\Delta\sin Az$, $v=\cos\Delta\cos Az$, $\Delta$ is the elevation angle, $Az$ is the azimuth), $u_i$ and $v_i$ are the direction cosines corresponding to the beam direction on the object under study. It follows from Eq. (\ref{eq2}) that at 40 MHz, the power pattern width of the sub-array is about 20.4$^\circ$. The experimental value of the power pattern width can be found from the interferometric response of radio sources Cyg A and Cas A. Times of passage of the power pattern at the level of half power are $\Delta t_{\rm Cyg\ A}$ = 110$\pm$2 min and $\Delta t_{\rm Cas\ A}$ = 162$\pm$2 min. The pattern width in degree is found from the following expression
\begin{displaymath}
\Theta\approx(360/(24 \cdot 60))\Delta t\cos\delta\,,
\end{displaymath}
where $\Delta t$ is taken in minutes, and the source declinations are $\delta_{\rm Cyg\ A}\approx 40.7339^\circ$ and $\delta_{\rm Cas\ A}\approx 58.5483^\circ$. Consequently, we have $\Theta_{\rm Cyg\ A}\approx20.9^\circ\pm1^\circ$ and $\Theta_{\rm Cas\ A}\approx21.2^\circ\pm1^\circ$ which are in good agreement with the theoretical calculations. Note that the beamwidth of our interferometer is about three times smaller. Therefore, our flux density measurements are much less contaminated by the extended Galactic background emission.

Relative sensitivity of radio-source measurements is defined as
\begin{displaymath}
\frac{\Delta S_{\nu,\rm min}}{S_{\nu,0}}=\frac{2kT}{S_{\nu,0}\,A_{eff}\sqrt{\Delta\nu\Delta\tau}}\,,
\end{displaymath}
where $S_{\nu,0}$ is the flux density of a source under observations, $k$ the Boltzmann constant, $\Delta\nu$ = 4kHz and $\Delta\tau$ = 25 sec are the frequency and the time resolutions, $T$ denotes the brightness temperature of the Galactic background. At 40 MHz the flux density of Cas A is about 27000 Jy (1Jy = 10$^{-26}$ W m$^{-2}$ Hz$^{-1}$), whereas the brightness temperature of the Galactic background in this direction of sky equals 15000 K. From here it follows $\Delta S_{\nu,\rm min}/S_{\nu,0}\approx 1.4\times 10^{-2}$. For the interferometer, this value is $\sqrt{2}$ times smaller, that is $\approx 10^{-2}$. Thus, despite the relatively small effective area of the GURT sub-array, several powerful cosmic radio sources can be observed with sufficiently high sensitivity. This produces highly reliable data in good quality. For this purpose the Cas A is just an appropriate radio source.

\subsection{Observations}

Our observations lasted from May 17 to October 11, 2019. They included records of radio emission with Cas A and Cyg A separately. To provide the flux-density measurement of Cas A, we used the Cyg A as a reference source with the well-known flux density which can be considered practically unchanged in comparison with Cas A. The ratio of radiation flux densities of the Cas A and Cyg A sources is related to the ratio of the measured signal powers in the direction of these sources as follows
\begin{equation}
r = K\frac{P_{\rm Cas\ A}}{P_{\rm Cyg\ A}}\,.\label{eq3}
\end{equation}
Here $K$ is the correction factor that depends on the angular dimensions of the sources, the absorption of radio radiation in the ionosphere due to differences in the angles of the location of the sources, differences in the brightness temperature of the Galactic background near Cas A and Cyg A, flickering of the sources on the inhomogeneities of the ionosphere, and the dependence of the antenna gain on different reception directions. The use of interferometers with a small base makes it possible not to take into account the difference in the angular sizes of the sources, if they are significantly smaller than the pattern of the interferometer. However, the base cannot be too small, as this widens the interferometer beam, which can capture the Galactic background around the sources differently. Our choice of base length was determined by the GURT array configuration. Howbeit the contamination of accurate flux density measurements by the Galactic background emission is much less than by other factors affecting the value of $K$. This is supported by the fact that if we consider on the accurate all-sky map of \citet{Dowell2017} at 35 MHz the positions of the beam of our interferometer at the locations of Cas A and Cyg A, then it is clear that for both of them this beam captures, on average, approximately the same intensity region of the Galactic plane. At higher frequencies the situation is even better. Let us also recall about the favorable geographic location of the GURT radio telescope which allows observing the sources of Cas A and Cyg A at almost the same zenith angle of $\approx9^\circ$ at the time of their upper culminations. Thus, the parameters of the phased antenna array in the directions to the sources may become almost identical. The significant factor that prevents the adjustment of the correction factor $K$ to one is the ionospheric flickering.  To reduce the influence of the ionosphere on the ratio of source fluxes, our observations were made in quiet conditions, and in further processing only the sessions obtained in conditions of a relatively undisturbed state of ionosphere were used. In addition, the results were averaged over a number of sessions. The number of favorable sessions was 58. Eventually the mean error of our measurements is about 10--15$\%$.

\begin{figure}
\centerline{\includegraphics[width=1.0\hsize]{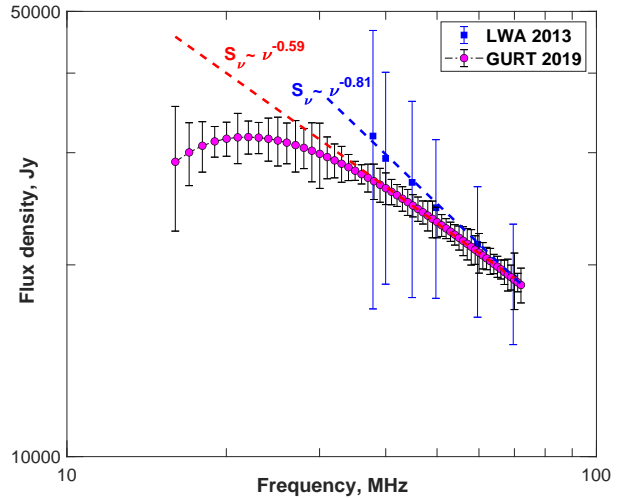}}
\caption{\label{fig5_CasA_flux}
The radio spectrum of Cas A in the range of 16--72 MHz. Red and blue dash lines indicate spectral indexes of the measured radiation at 38--70 MHz for Cas A obtained with the GURT and the LWA \citep{Dowell2017} radio telescopes, respectively.}
\end{figure}

Cygnus A is the most powerful extragalactic source of radio emission ($\sim$10$^{38}$ W) in its constellation (as noticed by the letter A in the name) and one of the brightest in the entire sky. It was discovered by Grote Reber in 1939. It is distant from us at 600 million light years and has a redshift of $z = 0.057$. The significant difference in the radio emission power of Cyg A and the closest external star system to us such as the Andromeda galaxy ($\sim$10$^{32}$ W) led later to the division of galaxies into two types one of which includes normal galaxies, similar to the Andromeda Nebula, whereas another consists of radio galaxies. In the meter and decameter wavelength ranges, the flux density of Cyg A, expressed in Jy, is described by the following relation
\begin{eqnarray}
S_{\nu,\ \rm Cyg\ A}&=&3.835\times10^5\left(\frac{\nu}{\rm MHz}\right)^{-0.718}\nonumber\\
&\times&\exp\left[-0.342\left(21.713/\left(\frac{\nu}{\rm MHz}\right)\right)^{2.1}\right]\,,\label{eq4}
\end{eqnarray}
where $\nu$ is the observation frequency in MHz \citep{Kassim1996,Vinyaikin2006,Men2008}. The exponential term of $S_{\nu,\ \rm Cyg\ A}$ was introduced to account for line-of-sight Galactic thermal absorption independent of the power-law synchrotron emission intrinsic to the source. Galactic thermal absorption is expected since Cyg A lies in the Galactic plane. Using the calculated values of the flux density of Cyg A, obtained from Eq. (\ref{eq4}) together with the experimental data for the ratios of radio fluxes from Cas A and Cyg A, the flux density of Cas A on a selected frequency grid was found (see Figure~\ref{fig5_CasA_flux} and next comments below). To obtain the most reliable results, the values of the source flux-density ratios at different frequencies were averaged by frequency to a selected frequency grid and by days.

It should be pointed out that both Cas A and Cyg A lie in the Galactic plane, and therefore both are subject to low frequency turnovers in their spectrum from extrinsic thermal absorption (not just Cas A). However, the absorption towards each is different because they are located in different regions along the Galactic plane. Fortunately, since the absorbed spectrum of Cyg A has been previously characterized in the GURT frequency range, we can still use it to calibrate our measurements of Cas A.

\section{Data analysis}
\subsection{Spectral measurements}

Radio records were implemented with the ADR (advanced digital receivers) which serves as a standard backend for the GURT observations \citep{Zakharenko2016}. Its time resolution was about 1 s in the frequency resolution of 19.073 kHz. Each ADR had two entries to measure cross-spectra. Their examples are presented in Appendix A. In the observations of Cas A and Cyg A one entry of the ADR was connected to one GURT sub-array, and the second entry was with another sub-array. The measurements of the flux-density ratio of Cas A and Cyg A have been made from 12 MHz to 73 MHz in 1 MHz steps. Unfortunately, at $<$15 MHz the GURT records of the radio sources are weighed down with intensive radio frequency interference (RFI). Therefore, our data analysis covers mainly the frequency range 16--72 MHz.

\subsection{Filtration of direction-dependent effects}

The ratio of measured flux densities between Cas A and Cyg A contains a quasi-periodic ``ripple''. It is caused by systematic oscillations in the beam properties as a function of frequency \citep{Popping2008}. There can be a number of reasons for this. The GURT array is located above a non-ideal reflecting ground. Moreover, the beam pattern may show variability or distortions due to mutual coupling between array elements, which are direction dependent. Note also that the interferometric baseline in our experiment is non-coplanar. The ionosphere can also introduce direction dependent effects although these are minimized on the relatively short baselines used in our observations \citep{Erickson1984}. To mitigate these complications, many different methods have been developed \citep{Bhatnagar2013,Smirnov2011}. They are quite time-consuming and aimed at obtaining high-quality resolved radio images from interferometric measurements. The methodology of our experiment, since our sources are extremely bright and unresolved, is simpler, allowing us to use simpler techniques for mitigating these direction-dependent effects. Assuming that the quasi-periodic oscillations in the ratio of flux densities of Cas A and Cyg A are mainly caused by instrumental effects, they were filtered with the help of the empirical mode decomposition (EMD) of \citet{Huang1998}. This makes it possible to separate the smooth component from the oscillating ones. 

\begin{figure}
\centering
\includegraphics[width=1.0\hsize]{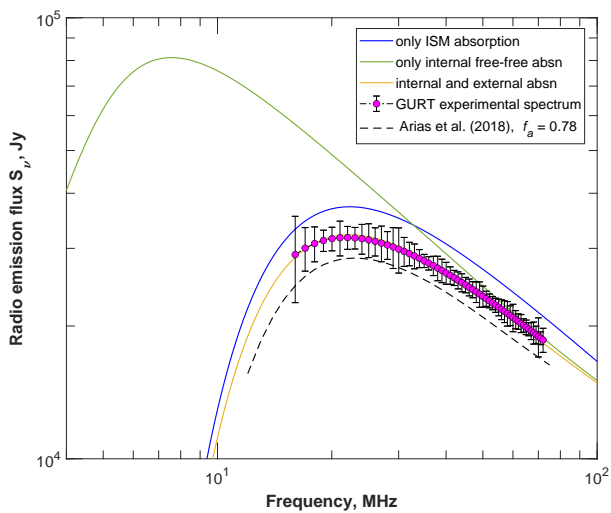}
\caption{\label{fig7_Comp} Effect of different forms of thermal absorption on the radio spectrum of Cas A as well as the fitting of the theoretical dependence to our experimental data obtained with the help of the GURT two-array interferometer ($f_a\approx$ 0.86). The observations cover the frequency range 16--72 MHz. The black dotted line below our measured spectrum shows the integrated spectrum of Cas A with the parameter values from \citet{Arias2018}, where $f_a=0.78$ is taken.}
\end{figure}

\subsection{Models of synchrotron source and free-free absorption}

Without loss of generality we start with the model of a homogeneous cylindrical synchrotron source \citep[see p.~97]{Pacholczyk1970}. Its functional form  is written as
\begin{equation}
J_\gamma(z)\sim z^{2.5}\left(1-\exp\left(-z^{-2-\gamma/2}\right)\right)\,,\label{eq5}
\end{equation}
where $\gamma$ characterizes a power-law distribution of electrons ($N(E)=N_0E^{-\gamma}$), and the dimensionless variable $z = \nu/\nu_1$ is the normalized frequency. In fact, the flux density of the radio source $S_\nu=AJ_\gamma(\nu/\nu_1)$ depends on three parameters: $\gamma$, $\nu_1$ and $A$. They describe the synchrotron emission from the SNR, the structure and features of which are known roughly. Both value of $\nu_1$ and $A$ are dependent on the magnetic-field strength $B$. Following the results of \cite{Arias2018}, we adopt $B=$ 0.78 mG. Next, using the experimental measurements of the Cas A flux density at the frequency of 1~GHz, taken from the aforesaid reference, we come to $\gamma=2.52$, $\nu_1=6.15$~MHz and $A=140500$~Jy. Thus, the above expression gives the unabsorbed synchrotron spectrum of Cas A.

The synchrotron emission propagates to observers through the SNR and the ISM, where it is absorbed by intervening thermal gas. The absorption consists of two components. One of them (internal) occurs inside the SNR, more specifically in its unshocked ejecta usually difficult to study. Another (external) is caused by the local ISM located around the SNR, where the remnant shock front has not yet reached. Following \citet{Kassim1995} and \citet{DeLaney2014}, there is internal free-free absorption in Cas A due to the unshocked ejecta . The medium causing internal absorption is interior to the bright radio shell, so it does not absorb radio emission generated in the front part of the shell. For external absorption this effect does not take place. Hence, the flux density can be modeled as
\begin{equation}
S_\nu = \left(S_{\nu,\rm front}+S_{\nu,\rm back}e^{-\tau_{\nu,\rm int}}\right)e^{-\tau_{\nu,\rm ISM}}\,,\label{eq6}
\end{equation} 
where $\tau_{\nu,\rm int}$ and $\tau_{\nu,\rm ISM}$ are the free-free optical depth for the unshocked ejecta and the ISM, respectively \citep{Arias2018}. As Cas A is quite clumpy, the relative synchrotron brightness contributed from the front and back halves of the remnant, as viewed from Earth, differ. Taking $S_{\nu,\rm front}=f_aS_\nu$ and $S_{\nu,\rm back}=(1-f_a)S_\nu$, we consider the absorption on the two halves of the shell. The free-free optical depth is determined in the Rayleigh–Jeans approximation \citep[see pp. 250-252]{Wilson2009}. Then the value reads
\begin{equation}
\tau_\nu = 3.014\times 10^4\,Z\left(\frac{T}{\rm K}\right)^{-3/2}\left(\frac{\nu}{\rm MHz}\right)^{-2}\left(\frac{EM}{\rm pc\ cm^{-6}}\right)g_{\rm ff}\,,\label{eq7}
\end{equation}
where $EM$ is the emission measure, $Z$ the average number of ion charges, $T$ the electron temperature of the absorbing medium, and $g_{\rm ff}$ denotes the Gaunt factor given by
\begin{equation}
g_{\rm ff}=\left\{
\begin{array}{ll}
\ln\left[49.55\,Z^{-1}\left(\frac{\nu}{\rm MHz}\right)^{-1}\right]+1.5\ln\frac{T}{\rm K} &  \\
1 \qquad \textrm{for}\qquad\frac{\nu}{\rm MHz}\gg\left(\frac{T}{\rm K}\right)^{3/2}\,.  & 
\end{array}\right.\label{eq8}
\end{equation}
The emission measure is defined along the line-of-sight as
\begin{equation}
\frac{EM}{\rm pc\ cm^{-6}}=\int_0^\frac{s}{\rm pc}\left(\frac{n_e}{\rm cm^{-3}}\right)\,\left(\frac{n_i}{\rm cm^{-3}}\right)\,d\left(\frac{s}{\rm pc}\right)\label{eq9}
\end{equation}
in terms of electron density $n_e$ and ion density $n_i$. Due to the free-free absorption, the integrated spectrum of Cas A peaks in the GURT observing band, as indicated by our radio measurements shown in Figures~\ref{fig5_CasA_flux} and \ref{fig7_Comp}. In the epoch of 2019 the spectrum peaks at 22 $\pm$ 0.5 MHz, and its value is 31743 $\pm$ 1894~Jy. 

\subsection{Measurements of free-free absorption parameters for Cas A}

From \citet{Arias2018} the ISM absorption along the line of sight to Cas A has $EM$ = 0.13 pc cm$^{-6}$ for a 20 K ISM with $Z=1$, whereas the internal absorption is characterized by a temperature of $T$ = 100 K and an average ionization state of $Z$ = 3 with an average emission measure of $EM$ = 37.4 pc cm$^{-6}$. In this case the value $f_a=0.78$ of the radio emission emerging from the projected area of the reverse shock comes from the foreground side of the shell. Consequently, one can draw the flux density spectrum of Cas A in the presence and absence of absorption. Note that the measurement of $EM$ values due to ISM absorption gave only an upper limit on the average electron density along the line of sight. On the other hand, the internal free-free absorption was measured from emission measure maps having extreme values. Pixel blanking up to 40\% was employed to exclude extreme values, nonetheless the derived $EM$ = 37.4 pc cm$^{-6}$ may reflect a residual systematic error. The symbol $Z$ is the average atomic number of the ions dominating the absorbing regions. In particular, the value of $Z_{\rm int}=$ 2--3 in the unshocked ejecta of Cas A is more reasonable, as it is the average number of ionizations. Following \citet{Arias2018}, there is some reason to believe that the temperature of the ISM and inside the unshocked ejecta of Cas A is 20 K and 100 K, respectively. However, they are based on assumptions, although the $T = 100$ K assumption is supported by \citet{Raymond2018}. Overall, given the uncertainties, independent measurements of the parameter values for the internal and external absorption in Cas A as presented in \citet{Arias2018} are useful.

\begin{table*}
\caption{Parameters of absorbing regions in Cas A obtained from the integrated continuum flux-density spectrum measured with the GURT interferometer.}%
\label{tab1}
\centering
\small
\begin{tabular}{cccccccc}
\hline\hline
 &  &  &  &  & &  & \\
\textbf{Parameters} & $EM_{\rm int}$ (pc cm$^{-6}$)  & $EM_{\rm ISM}$ (pc cm$^{-6}$) & $f_a$ & $T_{\rm int}$ (K) & $Z_{\rm int}$ & $T_{\rm ISM}$ (K) & $Z_{\rm ISM}$ \\
 &  &  &  &  & &  & \\
\hline
 & & & & & & & \\
\textbf{Values} & 37.36 $\mp ^{0.03} _{0.02}$ & 0.17 $\pm ^{0.11} _{0.03}$ & 0.86 $\pm ^{0.04} _{0.03}$ & 100.02 $\pm ^{0.01} _{0.01}$ & 2.55 $\mp ^{0.27} _{0.22}$ & 20.04 $\pm ^{0.01} _{0.02}$ & 0.51 $\mp ^{0.29} _{0.23}$ \\
 & & & & & & & \\
\hline
\end{tabular}
\end{table*}

Assuming values of $EM$, $T$, $Z$ and $f_a$ from \citet{Arias2018}, there is a noticeable discrepancy between predictions of Eq. (\ref{eq6}) and the flux-density spectrum of Cas A obtained in our observations. To illustrate this discrepency, we derived the specified values based on our experimental data. Note that the optical depth $\tau_\nu$ is linear with respect to the emission measure $EM$, whereas $T$ and $Z$ give a nonlinear contribution. To fit our experimental spectrum to the theoretical curve $S_\nu$ under absorption, we just need to set up seven parameters. For this purpose we used the nonlinear curve-fitting in least-squares sense. The parameter values found are presented in Table~\ref{tab1}. $EM_{\rm int}$, $T_{\rm int}$ and $T_{\rm ISM}$ agree closely with those in \citet{Arias2018}, but the value of $EM_{\rm ISM}$ is $\sim 27\%$ higher, and $f_a$ is $\sim 10\%$ higher, i.\,e. $f_a=$ 0.86 instead of 0.78. The latter difference is probably due to the fact that we cannot resolve the unshocked ejecta against the source, as we do not make exactly the same measurement. Our value of $f_a$ implies that in the absorbing region most of the radio emission is dominated by the front side of the shell. \citet{Arias2018} also find that most of the emission is dominated from the front side of the shell. So we are in agreement. The average atomic numbers of the ions in the internal and external absorption regions are $Z_{\rm int}\approx 2.55$ and $Z_{\rm ISM}\approx 0.51$. The errors listed in Table 1 were derived from errors of the measured spectrum. The comparison between theoretical and experimental results is shown in Figure~\ref{fig7_Comp}. Using the fitted spectrum, we find its maximum 32191 Jy at 22.345 MHz.

\begin{figure}
\centerline{\includegraphics[width=1.0\hsize]{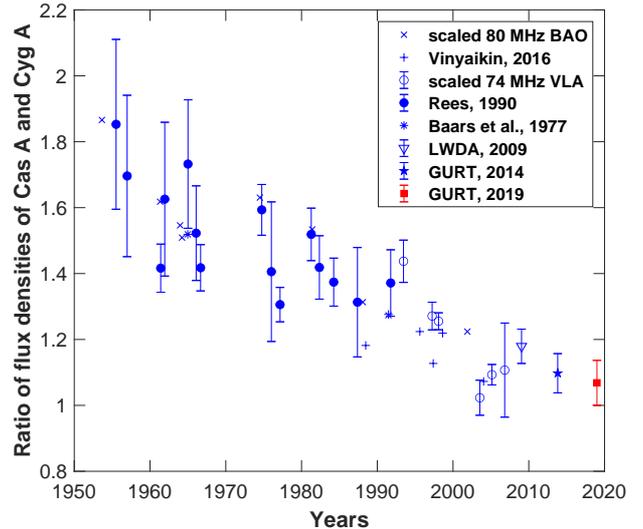}}
\caption{\label{fig6_38MHz}
Results of many different measurements for the flux-density ratio of Cas A and Cyg A at 38 MHz in time (from 1954 to 2019) that were published in literature \citep{Baars1977,Rees1990,Vinyaikin2006,Helmboldtl2009}. Also included is scaled versions of the 80 MHz data of \citet[BAO]{Martirossyan2002} as well as the GURT observations in 2014 \citep{Bubnov2014} and 2019.}
\end{figure}

\section{Secular decline of the radio flux of Cas A}
The most important characteristic of radiation from any radio source is its spectrum. In radio astronomy, the spectral flux density from nonthermal sources is usually defined as $S_\nu = C\,\nu^{-\alpha}$, where $\nu$ is the frequency of radio emission, $C$ a constant value, and $\alpha$ denotes the spectral index. In a very wide range of frequencies, the same value of the parameter $\alpha$ rarely succeeds in representing the change in $S_\nu$, i.\,e. the spectra show curvature. For example, in many metagalactic radio sources, the spectral index in the decimeter and centimeter ranges is greater than in the meter range. Sometimes such radio sources have a smooth increase in $\alpha$ with increasing frequency. In 2013 the measurements of the Cas A spectral index gave the value of 0.81 at 38--70 MHz \citep{Dowell2017}. Our results in 2019 lead to $\alpha=0.59\pm 0.02$ in the same frequency range (see Figure~\ref{fig5_CasA_flux}). This difference in the values of the spectral index from 2013 to 2019 reflects a time- and frequency-varying secular decline that has been recognized for some time in Cas A. This secular decline is also detected in the flux-density ratio of Cas A and Cyg A at 38 MHz shown in Figure~\ref{fig6_38MHz} for many years of observations. As can be seen, the radio data obtained by us are consistent with the expected decrease in the flux density of Cas A. The declining Cas A flux density is a natural consequence of adiabatic cooling, originating from the SNR expansion. Considering the still enormous magnitude of the radio flux of Cas A, sensitive observations, even separated from each other by small time intervals, reveal not only the age-related decrease of the flux, but also a change in the spectrum. 

\begin{center}
\begin{table}[t]%
\centering
\caption{Rate of decrease of the flux density for the radio emission of Cas A in dependence of frequency from our observations of 2019 and according to the data of \citet{Braude1969} and \citet{Baars1977}.\label{tab2}}%
\tabcolsep=0pt%
\begin{tabular*}{20pc}{@{\extracolsep\fill}lccccc@{\extracolsep\fill}}
\hline\hline
\textbf{Frequency,} & \textbf{$S_{\nu,1966}$, Jy}  &  & \textbf{$S_{\nu,2019}$, Jy}  & \textbf{Reduction of $S_\nu$}  \\
\textbf{MHz} &  &  &  & \textbf{per year, \%}  \\
\hline
14.7 & 65000 $\pm$ 9100 & \quad\qquad & 14881 $\pm$ 6434 & 2.92 $\pm$ 1.07 \\
16.7 & 60000 $\pm$ 8400 & \quad\qquad & 30291 $\pm$ 3400 & 1.27 $\pm$ 0.34 \\
20 & 65000 $\pm$ 9100 & \quad\qquad & 31838 $\pm$ 1940 & 1.32 $\pm$ 0.29 \\
22.25 & 51634 $\pm$ 2570 & \quad\qquad & 31934 $\pm$ 1894 & 0.90 $\pm$ 0.15 \\
25 & 58000 $\pm$ 8120 & \quad\qquad & 31588 $\pm$ 2759 & 1.13 $\pm$ 0.31 \\
26.3 & 45288 $\pm$ 2028 & \quad\qquad & 31256 $\pm$ 2288 & 0.70 $\pm$ 0.16 \\
38 & 36380  $\pm$ 1339 & \quad\qquad & 27158 $\pm$ 1728 & 0.55 $\pm$ 0.14 \\
\hline
\end{tabular*}
\end{table}
\end{center}

The rate of decrease in the flux density of Cas A for the period 1966--2019 was calculated by using the following expression
\begin{equation}
S_{\nu,2019}=S_{\nu,1966}\left(1-\frac{p}{100}\right)^n\,,\label{eq10}
\end{equation}
where $S_{\nu,2019}$ and $S_{\nu,1966}$ are the flux densities in 2019 and 1966, respectively. Here $p$ is the annual decrease of flux density in percent, and $n$ denotes the observation period in years. The results of calculations according to Eq. (\ref{eq10}) are summarized in Table~\ref{tab2}. The flux-density values of Cas A in the 1966 epoch at frequencies of 14.7 MHz, 16.7 MHz, 20 MHz and 25 MHz were obtained by \citet{Braude1969} using the Ukrainian T-shape Radiotelescope, first modification (UTR-1). The flux-density data at 38 MHz in the 1966 epoch were taken from \citet{Baars1977}. Based on Table~\ref{tab2}, it should be noticed that the decrease in the Cas A flux density from 1966 to 2019 per year in the 15--38 MHz frequency band is 1.26 $\pm$ 0.35$\%$ per year on average. At the frequency of 38 MHz, this decrease tends to 0.55 $\pm$ 0.14$\%$.

\section{Discussion}

\subsection{Internal electron density and mass}
The measured magnitudes of an emission measure $EM$, a temperature $T$, and an average number of charges of the ions $Z$ for internal free-free absorption allow us to estimate the electron density and the mass of the unshocked ejecta. However, the estimates are sensitive to the ejecta geometry which is not exactly known, and its description requires assumptions. If the electron density in the unshocked ejecta is constant, and $n_e=n_i$, then Eq.(\ref{eq9}) simplifies to $n_e=\sqrt{\frac{EM}{l}}$, where $l$ is the combined thickness of the unshocked ejecta sheets. The clumping of the unshocked ejecta can reduce the total volume of the medium, increasing its density. By the clumping factor $0<C_f\leq1$ the effect is taken into account. Then the electron density in the unshocked ejecta is written as $n_e=\sqrt{\frac{EM}{lC_f}}$. If $C_f=1$, there is no clumping. For this effect to be, $C_f < 1$ indicates a degree of clumping. As \citet{DeLaney2014}, we take these sheets to be 0.16 pc thick. For $C_f=1$ the internal electron density is about 15.3 cm$^{-3}$. If $C_f=0.67$, the value of $n_e$ increases to 18.7 cm$^{-3}$. 

The total mass of unshocked ejecta has the simple form $M=V\rho$, i.\,e. the product of its volume $V$ and its density $\rho$. Following \citet{DeLaney2014}, the total volume of internal absorbing material is $V = \pi R^2l$ = 1.1 pc$^3$. If the effect of clumping is present, we simply modify $R$ and $l$ by the clumping factor $C_f$ (as $C_fl$ and $C_fR$). As the main contributors to the mass are the ions, then their density is $\rho=Am_p\frac{n_e}{Z}$, where $A$ is an average mass number, and $m_p$ is the mass of the proton. Thus, the total mass of the unshocked ejecta yields
\begin{eqnarray}
M = &0.96 \pm ^{0.11} _{0.08}M_\sun&\left(\frac{A}{16}\right)\left(\frac{V}{\rm 1.1\ pc^3}\right)\left(\frac{2.55}{Z}\right)\nonumber\\
&&\times\left(\frac{n_e}{\rm 18.7\ cm^{-3}}\right)\left(\frac{C_f}{0.67}\right)^{5/2}\,.\label{eq9a}
\end{eqnarray}
Therefore, our estimate of the mass in the unshocked ejecta with the clumping effect gives about $0.96\pm 0.11M_\sun$, whereas without clumping the mass reaches $2.61\pm 0.31M_\sun$ that is close to $2.95\pm 0.48M_\sun$ obtained by \citet{Arias2018}. Taking into account the clumping contribution, our value agrees with the independent estimate of $0.49 ^{+0.47} _{-0.24} M_\sun$ from \citet{Laming2020}. Recently the similar estimate of 0.6--0.8$M_\sun$ has been found by \citet{Priestley2022}.

Considering the unshocked ejecta without clumping, the ejecta mass can be overestimated. If the progenitor mass before the explosion is 4--6$M_\sun$ \citep{Young2006}, the mass in the unshocked ejecta is quite high. There are two ways to solve this puzzle. One of them is to assume that the unshocked ejecta is clumped. Then the mass estimate becomes significantly smaller and more appropriate. On the other hand, the progenitor of Cas A is thought
to have 15--25$M_\sun$ as a star that lost its hydrogen envelope to a binary interaction \citep{Chevalier2003}. In this case the mass estimate in the unshocked ejecta without clumping effects would be acceptable. Unfortunately, the progenitor mass of Cas A is not yet definitely known. Therefore, often both scenarios are considered.

\subsection{Turnover in the low-frequency spectrum of Cas A}

The integrated radio spectrum of Cas A has been studied for a long time \citep{Kassim1989,Kellermann2009}. One of its remarkable features is an extreme at low frequencies. Following Table 2 of \citet{Baars1977}, in the epoch of 1966 the spectral turnover is located below 38 MHz, between 14.7 MHz and 20 MHz \citep{Braude1969}. The fitting of the flux-density spectrum gives 18 MHz as a spectral maximum \citep{Vinyaikin2007}. However, direct measurements of this turnover frequency are rather complicated. If the frequencies of the extrema in the spectral flux density of radio sources are in the centimeter, decimeter, or short-wave part of the meter wavelength range, then such measurements are usually fairly easy. Observations at the lower frequencies, especially below 20 MHz, are very difficult since they are significantly susceptible to the RFI and influenced by the cosmic medium and solar supercorona. Moreover, the contribution of the Earth's ionosphere makes the regular ground-based measurements at frequencies below 20 MHz not often successful and reliable. Not surprisingly, there are different estimates of the frequency, where the spectral maximum of Cas A is located \citep{Vinyaikin1987,Kassim1995,Vinyaikin2014}. Recently, the opinion has dominated that the turnover in the integrated spectrum of Cas A is at 20 MHz \citep{Arias2018}. From our observations of 2019 the frequency is 22 $\pm$ 0.5 MHz. Due to small errors of the measurements, the fitting of the measured flux-density spectrum estimates the value of 22.345 MHz. This may indicate that the spectral maximum of Cas A is not fixed in time and frequency, and from 1966 to 2019 it changed its position by 4 MHz towards higher frequencies. In addition, the peak decreases in intensity. For half a century, the largest value in the spectrum of Cas A has decreased by about half in magnitude. The most significant reason for this evolution is most likely a decrease in the intensity of synchrotron radiation from the source as it expands and cools.

This forces us to look at the free-free absorption by the ISM ionized gas along the line of sight to the source differently. Direct and reverse shock waves of the SNR can rake up the absorbing material inside the space between them. This addition to the ISM gives an increase in the emission measure of external absorption. Consequently, the optical depth becomes larger, and the spectral maximum of Cas A shifts towards higher frequencies. If we assume that the frequency shift by 4 MHz occurs only due to the emission measure of external absorption, then the value of the latter would increase from $\sim$0.11 to $\sim$0.17 pc cm$^{-6}$. In this regard, it should be emphasized that further low-frequency observations of the integrated spectrum of Cas A would be useful for a more detailed understanding of how the internal and external absorption changes with time.

\section{Conclusions}
We have measured the integrated radio continuum spectrum of Cas A at 16--72 MHz in the epoch of 2019. Using a two-element correlation interferometer based on the GURT array, the radio galaxy Cyg A was used as a reference radio source whose known spectrum was assumed to be constant. Fitting the spectral model of Cas A with free-free absorption effects from the ISM and interior unshocked ejecta, the parameter values ($EM$, $T$, $Z$ and $f_a$) of internal and external absorption were found. Comparing our measurements with published results for the flux density of Cas A in a wide frequency band, the secular decline was detected and matches the expected. We summarize our results as follows:

\begin{enumerate}
      \item The measured spectrum of Cas A with the GURT interferometric observations is consistent with the theoretical one within the measurement errors. The spectrum of Cas A is determined by not only the synchrotron emission mechanism, but by thermal absorption inside the SNR and in the ISM.
			\item In the epoch of 2019 the peak of the Cas A flux-density spectrum is 31743 $\pm$ 1894 Jy at 22 $\pm$ 0.5 MHz.
			\item From our measurements the area, associated with unshocked, ionized ejecta, internal to the reverse shock, has a temperature $T\approx 100$ K, an average ionisation state of $Z\approx 2.55$ and an average emission measure of $EM\approx 37.36 \, \rm{pc} \, \rm{cm}^{-6}$. 
			\item We measure the ISM absorption along the line of sight to Cas A to be $EM\approx 0.17 \, \rm{pc} \, \rm{cm}^{-6}$ for a 20 K ISM. \citet{Arias2018} determined that 78\% of the emission emerging from the region projected against the unshocked ejecta came from the foreground side of Cas A. Our slightly higher value of 86\% is not unexpected since we do not have the angular resolution to isolate the region of emission against the region of unshocked ejecta.
			\item The measured free-free absorption parameters, mentioned by \citet{Arias2018}, are close enough to our experimental estimates for their values, although some of them are clarified.
      \item Our estimate of the internal electron density in the unshocked ejecta is about 15.3 cm$^{-3}$, whereas the ejecta mass is 2.61$\pm 0.31M_\sun$. This unshocked ejecta is not clumped. If the clumping factor is 0.67, for example, then the electron density is 18.7 cm$^{-3}$, and the mass decreases to 0.96$\pm 0.11M_\sun$.
			\item In the frequency range of < 70 MHz, the flux density of Cas A decreases faster at lower frequencies than at higher frequencies. This is due to the fact that the spectral index decreases in time. If in the epoch of 2013 its value is about 0.81 in the frequency range of 38--70 MHz, then in 2019 the spectral index fell to 0.59 $\pm$ 0.02, i.e. the radio spectrum of Cas A is flattening. The observed effect is most likely primarily due to a flattening of the radio synchrotron spectrum of Cas A itself \citep{Vinyaikin2014}.
      \item At the frequency of 38 MHz, the decrease in the Cas A flux density from 1966 to 2019 is 0.55 $\pm$ 0.14$\%$ per year.
			\item It should be noted that most of the values of the free-free absorption parameters towards Cas A are consistent with other recently published results in the literature \citep[and references therein]{Arias2018}. But we also found differences, such as a 4 MHz shift for the spectral maximum of Cas A towards higher frequencies from 1966 to 2019. Our results are important because they confirm previous results in a completely independent way with a completely different instrument. Our instrument is one of the few in the world capable of operating at such low frequencies, and this independent confirmation is very important.
\end{enumerate} 

\begin{acknowledgements}
The authors acknowledge the partial support of the National Academy of Sciences of Ukraine under Contracts No. 0122U002459, 0122U002460 and 0122U002537. Personally, we are grateful to M. Arias for her useful remarks and discussions. We also thank the anonymous referee for very valuable comments that contributed to improve the quality of the manuscript. 
\end{acknowledgements}



\bibliographystyle{aa} 

\begin{appendix}
\section{Interferometric responses of Cyg A and Cas A in the GURT sub-arrays}

In this appendix we present examples of interferometer measurements made by the two sub-arrays within GURT, as described in Section~\ref{section2_1}.
\setcounter{figure}{0}
\renewcommand{\figurename}{Fig.}
\renewcommand{\thefigure}{A.\arabic{figure}}
\begin{figure}
\centerline{\includegraphics[width=0.49\paperwidth]{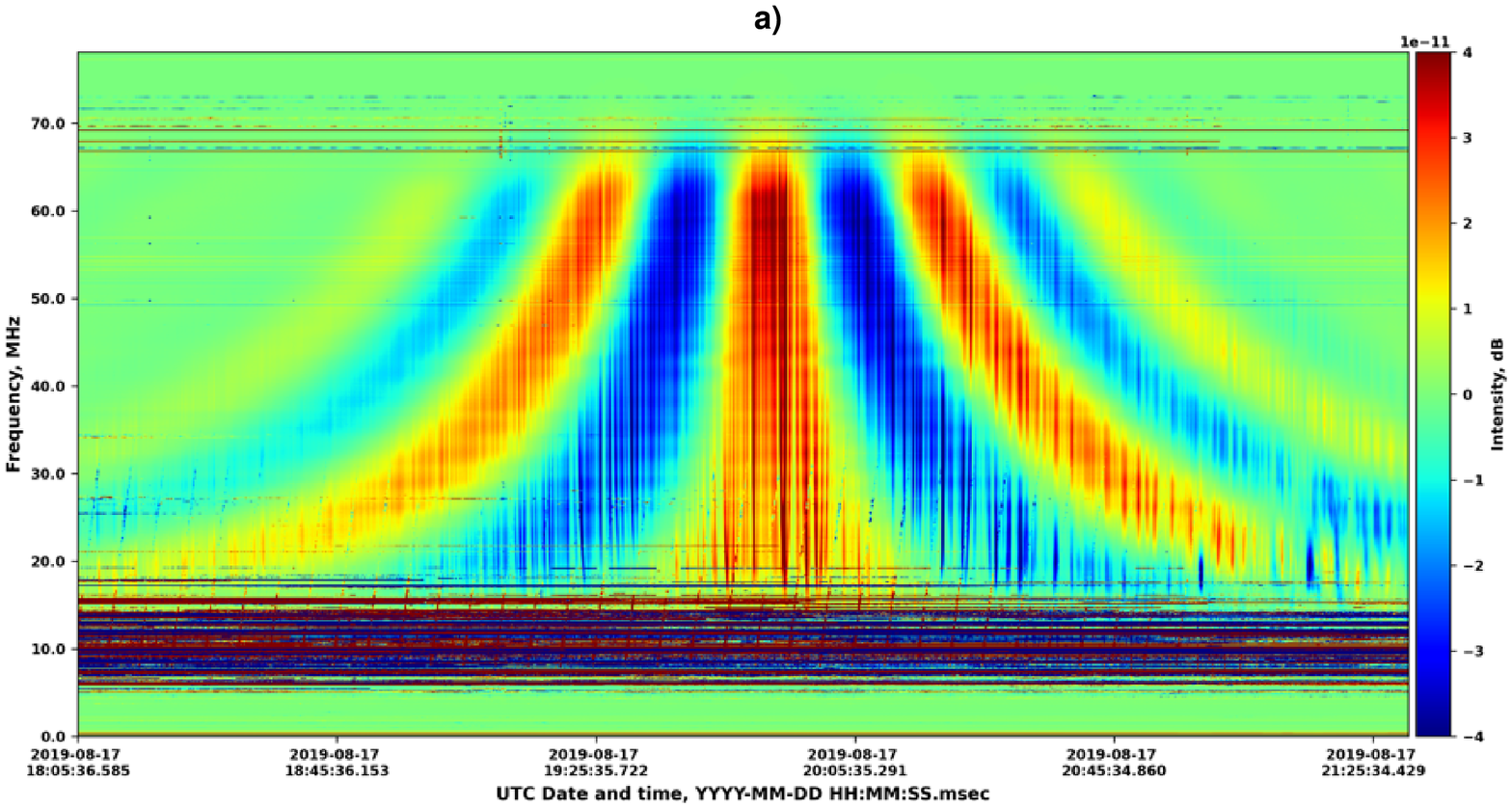}}
\centerline{\includegraphics[width=0.49\paperwidth]{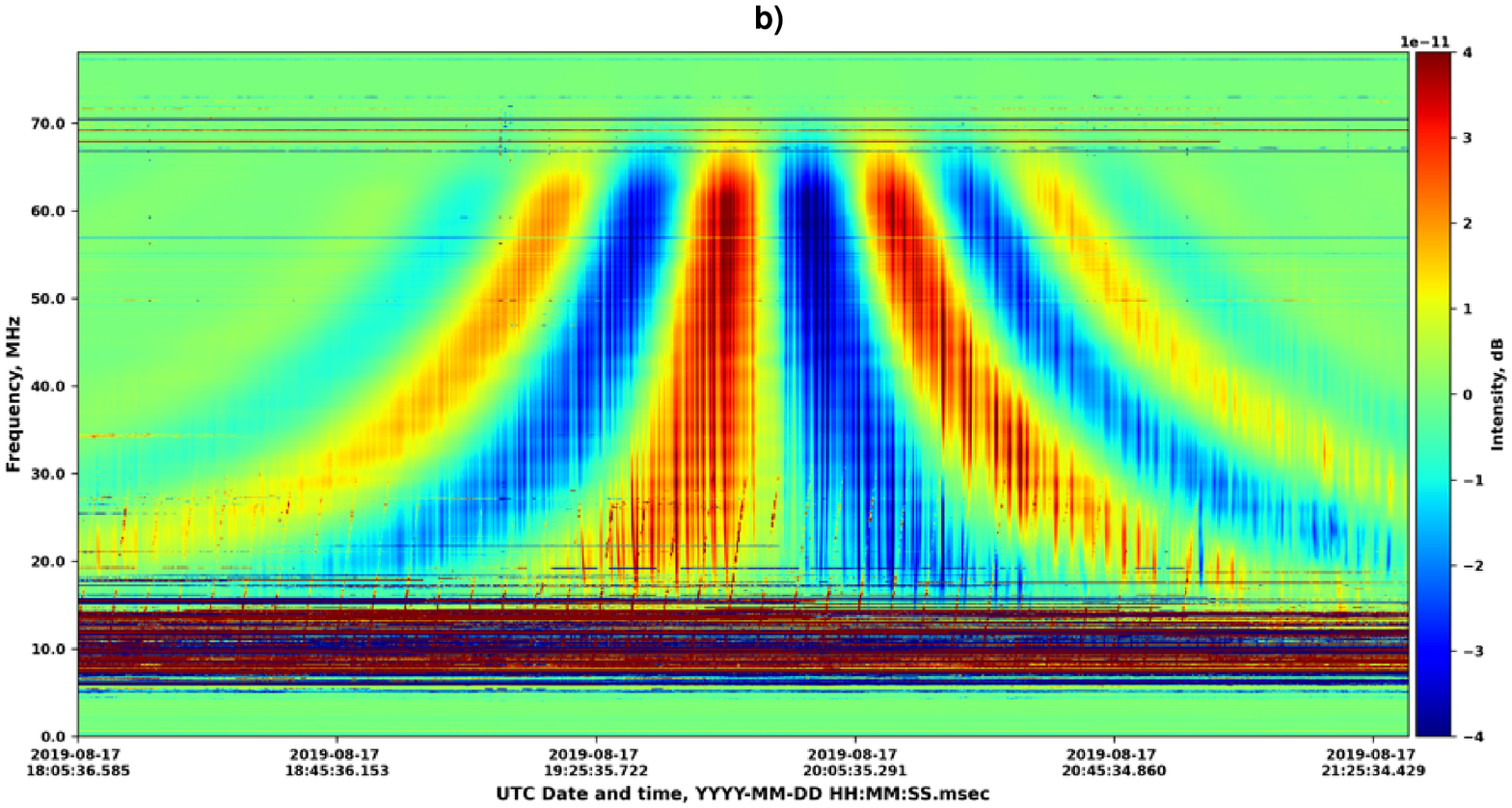}}
\caption{Dynamic cross-spectra of the interferometric response in the terms of even (a) and odd (b) parts from radio emission of Cyg A observed by two GURT sub-arrays on August 17, 2019. The radio source passes through the fixed patterns of antennas.}
\label{appfig1}
\end{figure}
\begin{figure}
\centerline{\includegraphics[width=0.49\paperwidth]{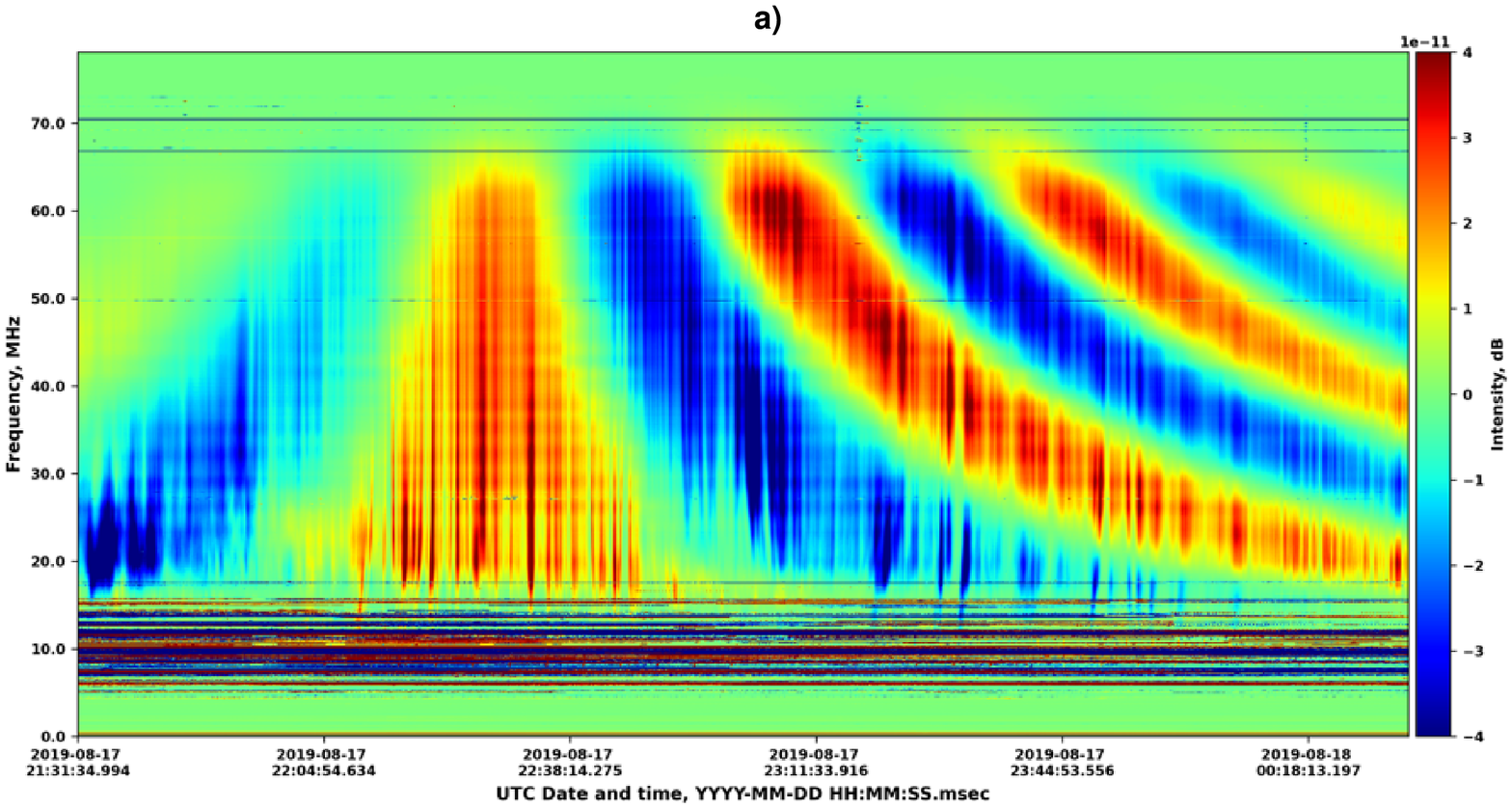}}
\centerline{\includegraphics[width=0.49\paperwidth]{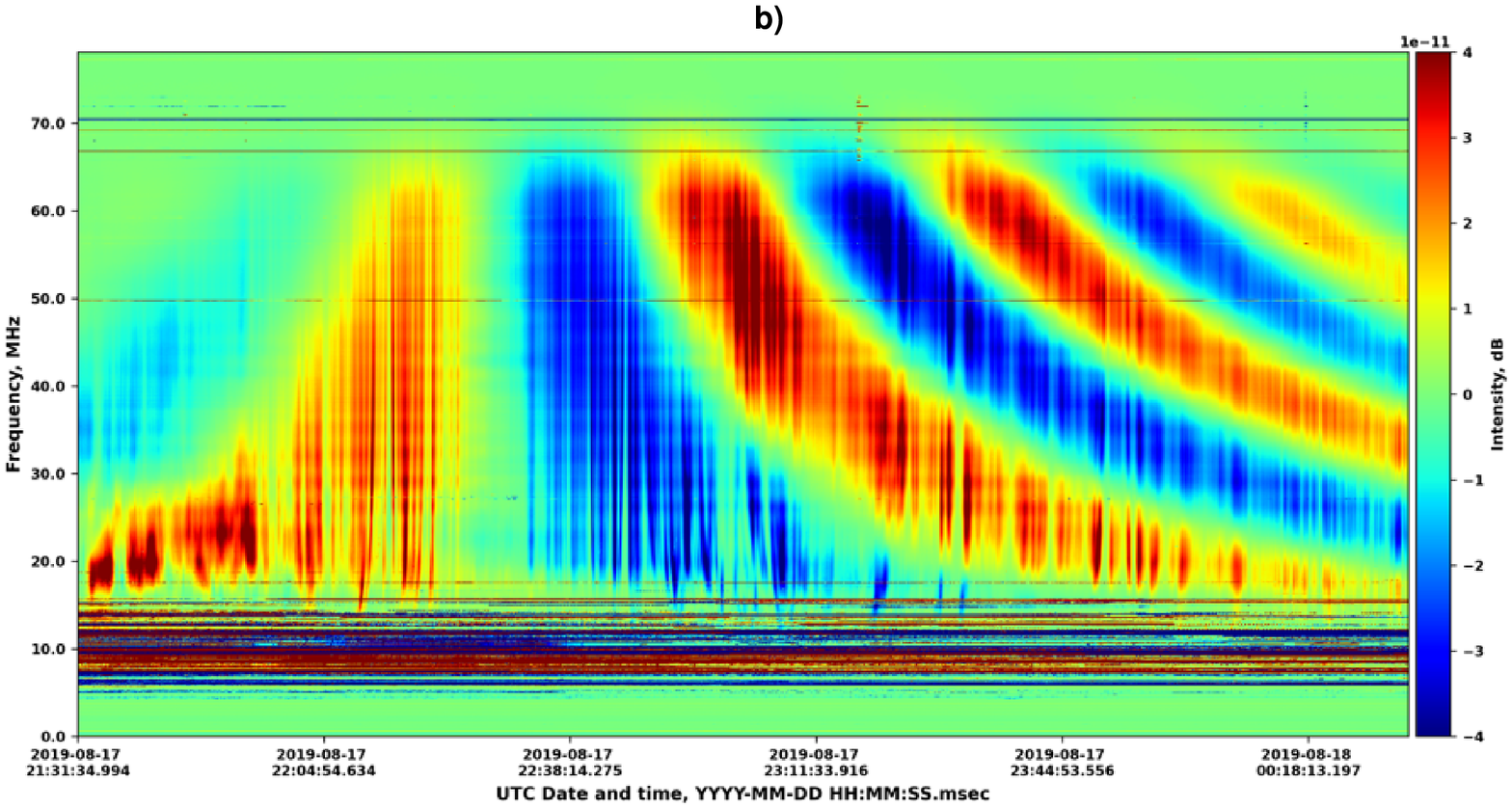}}
\caption{Dynamic cross-spectra of the interferometric response in the form of even (a) and odd (b) parts from radio emission of Cas A observed by two GURT sub-arrays on August 17, 2019. The recording of Cas A is limited at the beginning, since the time difference between the culminations of Cyg A and Cas A is about 3.5 hours, and Cyg A is observed earlier than Cas A. Therefore, the switching from one radio source to another occurs approximately in the middle of each observation session.}
\label{appfig2}
\end{figure}

\end{appendix}

\end{document}